\documentclass[conference,10pt]{IEEEtran}

\usepackage{amsmath,graphicx}

\usepackage{subfigure}
\usepackage{epsfig,graphicx,rotating}
\usepackage{amssymb,amsmath}
\usepackage{epstopdf}
\usepackage{tikz} 
\usepackage{float}
\usepackage{tikz-3dplot}
\usepackage{color}
\usepackage{pdfpages}
\usepackage{cite}      
\usepackage{booktabs}
\usepackage{psfrag}
\usepackage{xcolor}
\usepackage{url}

\makeatletter
\newcommand{\vast}{\bBigg@{4}}
\newcommand{\Vast}{\bBigg@{5}}
\makeatother

\usepackage{breqn}

\usepackage{tikz} 
\usetikzlibrary{decorations.markings,decorations.pathreplacing,decorations.pathmorphing,shapes,backgrounds,patterns,decorations.text}
\usepackage{tikz-3dplot}
\usepackage{color}
 \usepackage{pgfplots}

\title{Tutorial I: \\ Learning the Principles of Mobile Radio Propagation through Smartphone and CRFO\\
	}
\author{Prabhu Chandhar, Sathish Babu and Tamizh Elakkiya\\
Chandhar Research Labs, Chennai, India\\
Email: prabhu@chandhar-labs.com, sathish@chandhar-labs.com, tamizhjai@gmail.com\\ \\
Version: 1.0, \today
}

\usepackage{fancyhdr}
\begin{document}
%
\maketitle
\begin{abstract}
In this tutorial, we present three simple smartphone based experiments for understanding the basic concepts of mobile communications such as pathloss, Shadow fading, and small scale fading. We also explain the use of Collaborative Radio Frequency Observatory (CRFO), an online platform, for visualizing radio coverage maps. 
\end{abstract}


\section{Mobile Radio Propagation Modeling}
\begin{figure}[htbp!]
\centering
\includegraphics[trim={0 9cm 0 0},clip,width=\linewidth]{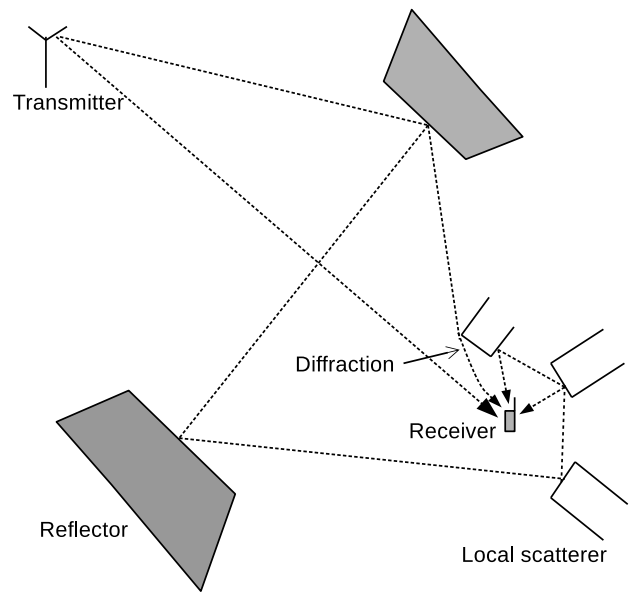}
\caption{Wireless propagation scenario}
\label{propagation_scenario}
\end{figure}

Channel modeling plays an important role in designing and analyzing mobile cellular communication systems. In wireless communications, the electromagnetic wave propagation is severely affected by the physical phenomena such as reflection, diffraction, and scattering \cite{Rappaport2001} as shown in Figure \ref{propagation_scenario}. Propagation models are used to characterize these propagation mechanisms and are categorized into large scale and small scale fading models. Propagation models that characterize the signal strength over large transmitter receiver separation are called large scale propagation models. Typically, the large scale propagation models are used to predict the coverage area of a transmitter. Propagation models that used to characterize the signal strength over short distances (a few wavelengths) or short time duration (a few milliseconds) are called small scale fading models. For an example, the distribution of envelope of small scale fading is modelled as Rayleigh and Rician distributions and the large-scale fading is modelled as Log-Normal distribution \cite{Rappaport2001,Stuber2001}. 

In this tutorial, we discuss the simple ways to learn the following mobile communication phenomenons:
\begin{itemize}
    \item Pathloss
    \item Shadow fading
    \item Small scale fading
\end{itemize}

\section{Required Software Tools}
The following tools are required to learn the fundamental propagation concepts of mobile communication systems.
\begin{enumerate}
    \item Collaborative Radio Frequency Observatory (CRFO)
    \item TRAI MySpeed App
    \item Shadow Lab App
\end{enumerate}

\subsection{CRFO}
CRFO is an online observatory containing radio frequency measurements from various locations in India \cite{crfo}. The primary purpose of CRFO is to enable collaborative research and development in the field of wireless communication in a cost-effective manner. The CRFO can be used for the following purposes:
\begin{itemize}
\item National spectrum monitoring in India
\item Dynamic Spectrum Access (DSA)
\item Cognitive Radio (CR)
\item Development of Machine Learning (ML) tools for wireless applications
\item Teaching advanced signal processing and wireless communications
\end{itemize}

\subsection{TRAI MySpeed App}
Telecom Regulatory Authority of India (TRAI) MySpeed App \cite{trai_myspeed} is a convenient tool to measure received signal strength, uplink data speed, downlink data speed and other network related information from the nearby 4G mobile towers.

\subsection{Shadow Lab App}
The Shadow Lab app \cite{shadow_lab} is developed by \textit{Chandhar Research Labs} \cite{crl} with the help of an open source android app `LTE Coverage Tool' developed by the National Institute of Standards and Technology (NIST), Public Safety Communications Research Division (PSCR), USA \cite{lte_cov_tool}.

The Shadow Lab app collects Reference Signal Received Power (RSRP) samples at regular time intervals (currently for every second) and plots three different graphs: histogram, PDF, and line chart. In the PDF graph, we can also see the mean and standard deviation of the collected RSRP samples.





\section{Experiment 1: Understanding Pathloss}
The objective of this experiment is to understand the pathloss phenomenon from the real-time data from the smart phones. The TRAI Myspeed app is used to collect RSRP samples from the received signals to create a .CSV file and to upload in CRFO web interface to visualize the RSRP heatmap.

\begin{figure*}[htbp!]
\centering
\subfigure[]{\includegraphics[scale=.115]{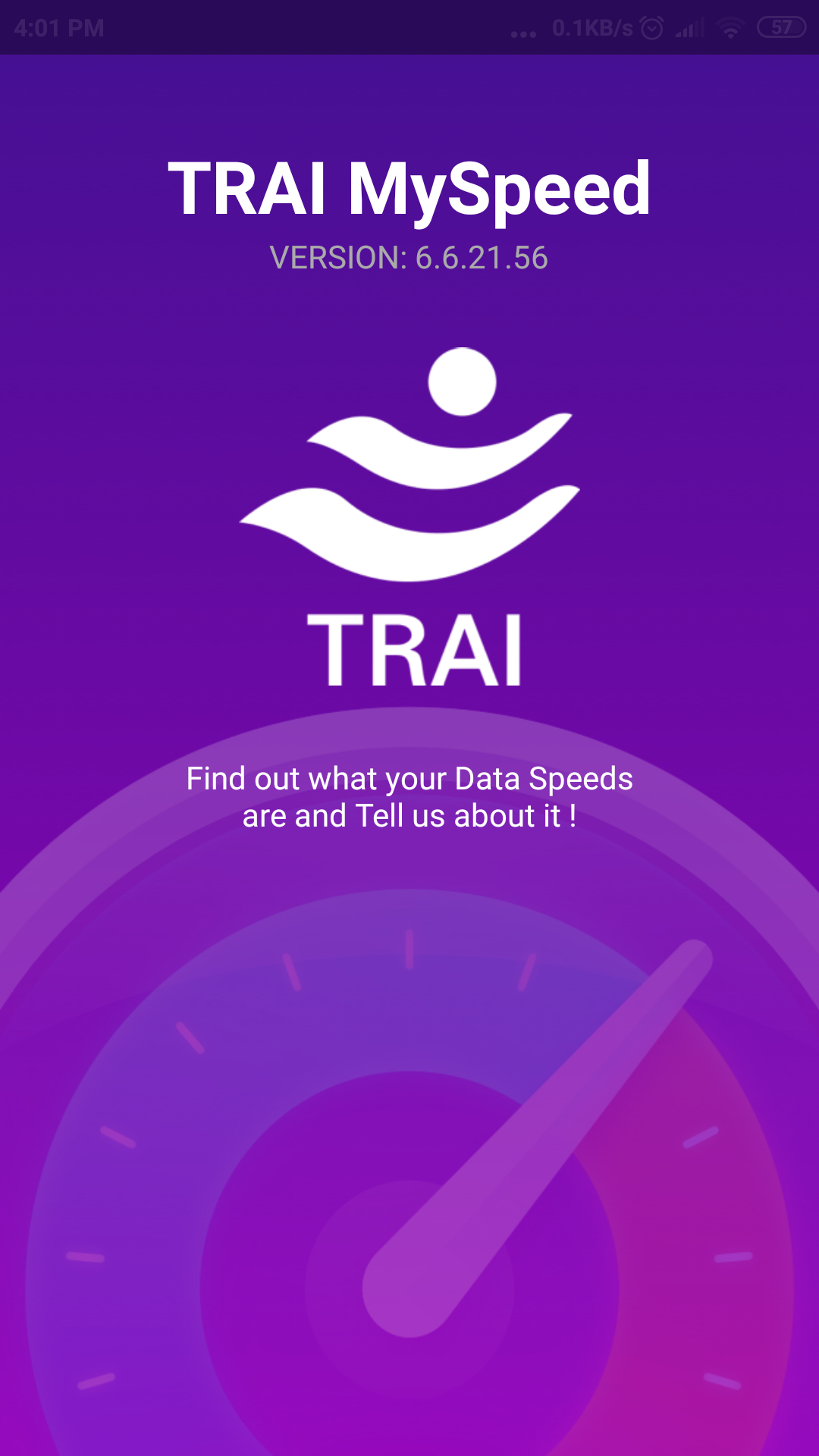}}
\subfigure[]{\includegraphics[scale=.115]{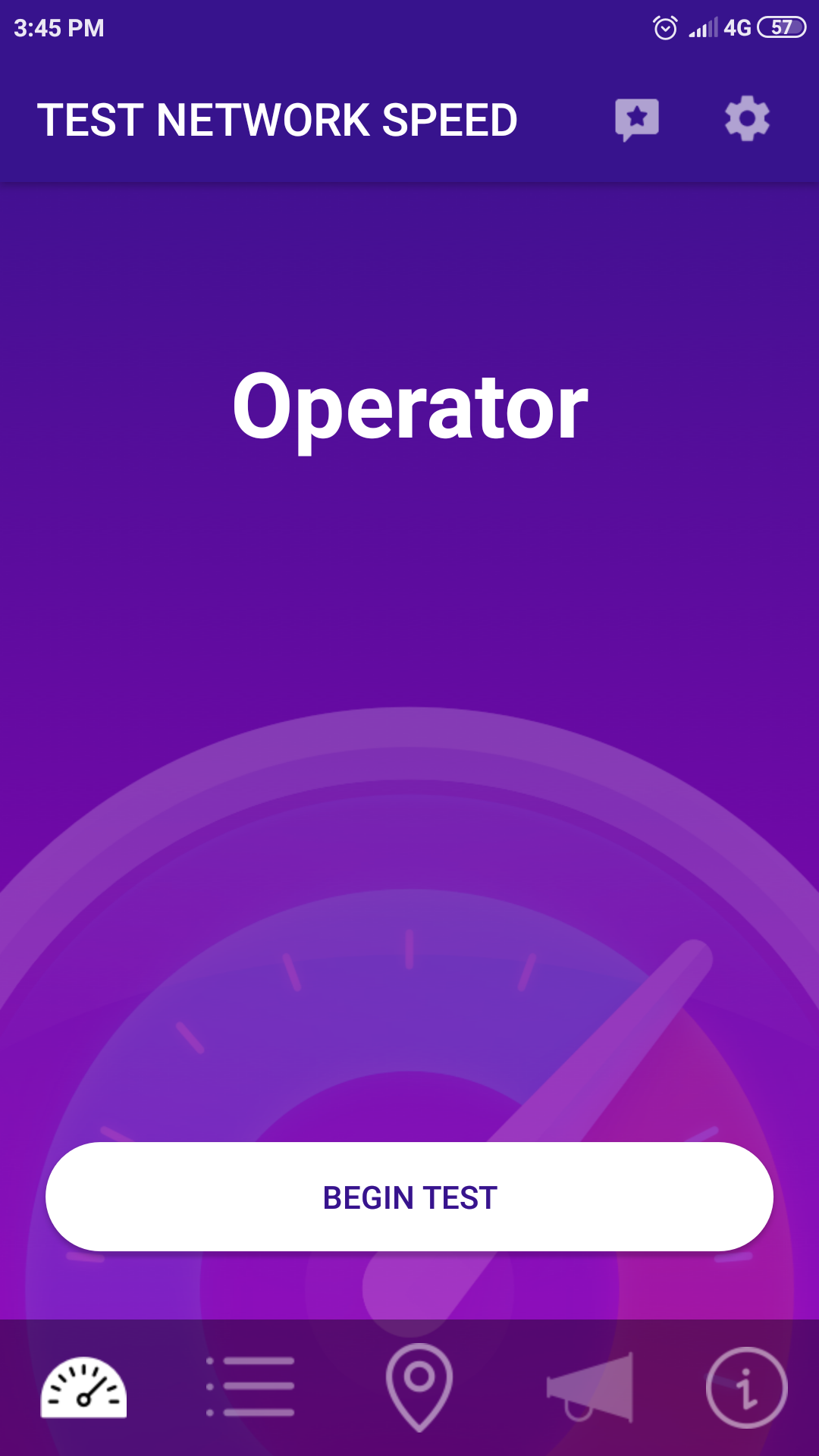}}
\subfigure[]{\includegraphics[scale=.115]{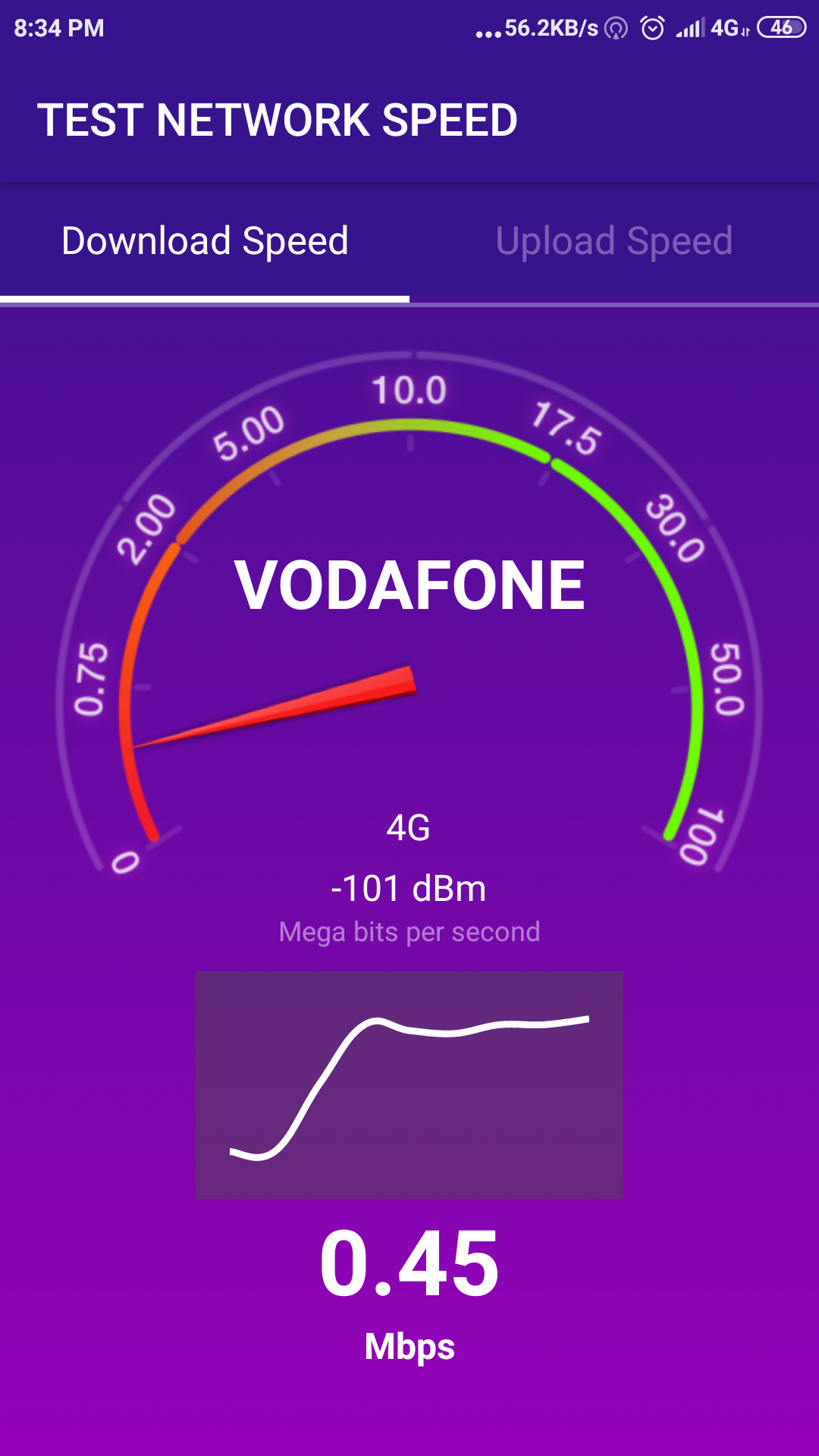}}
\subfigure[]{\includegraphics[scale=.115]{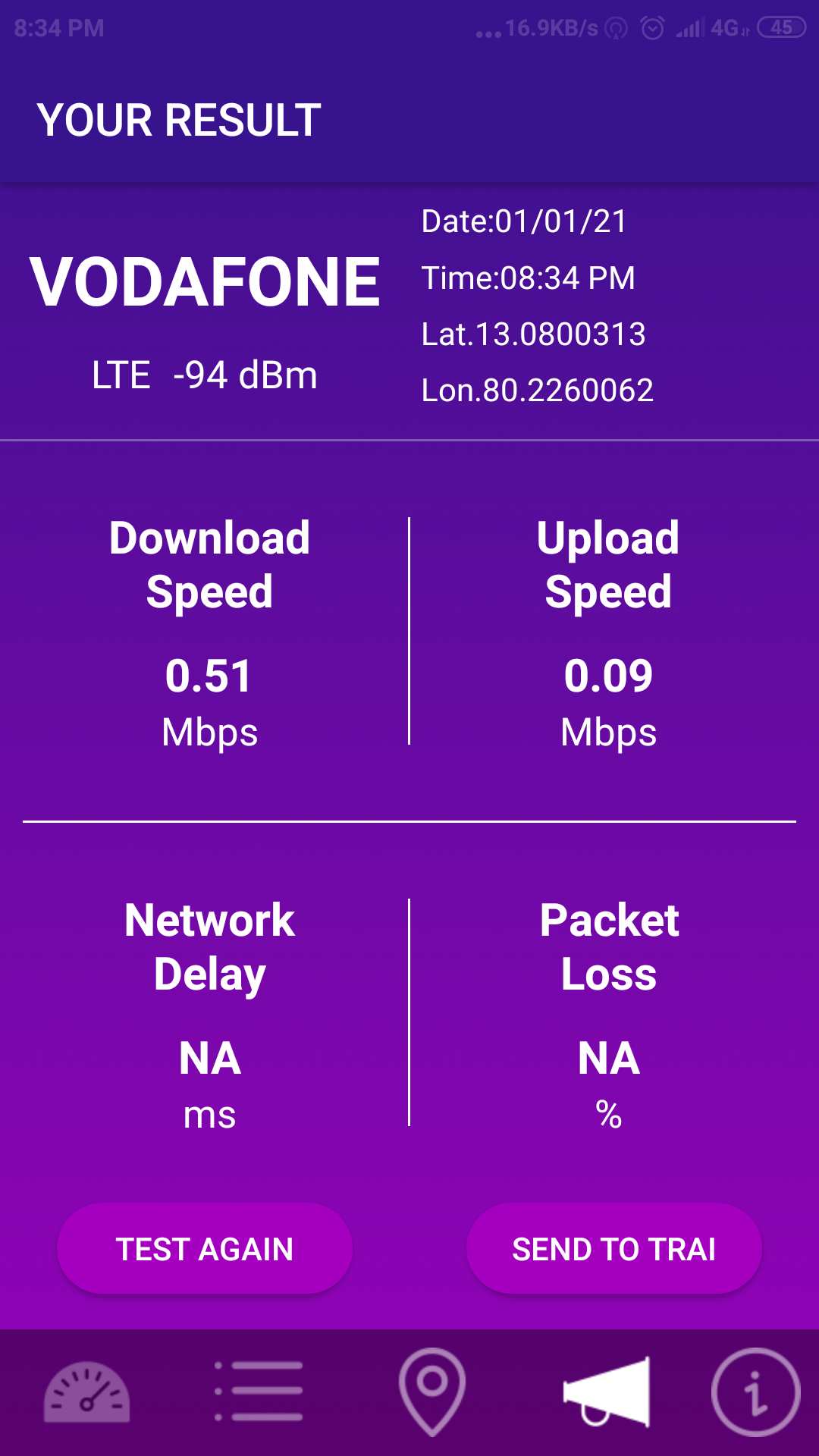}}\label{MySpeedTrai_result}
\caption{Screenshots of `TRAI MySpeed' app}
\label{TRAI app}
\end{figure*}

\subsection{Pathloss Model}
Pathloss represents the signal attenuation between a transmit and receive antenna as a function of propagation distance and other parameters. The average pathloss for a receiver located $d$ distance away from the transmitter is expressed by \cite{Rappaport2001}
\begin{equation}
\overline{PL}(d) = \bigg(\frac{d}{d_0}\bigg)^\alpha
\end{equation}
or 
\begin{equation}
\overline{PL}(d) = \overline{PL}(d_0) + 10 \alpha \log_{10}\bigg(\frac{d}{d_0}\bigg)   \ \ \ \text{dB},
\end{equation}
where $\alpha$ is pathloss exponent, $d_0$ is the close-in reference distance. $d_0$ is usually greater than the Fraunhofer distance $d_f$ which is given by \cite{Rappaport2001}
\begin{equation}
d_f = \frac{2D^2}{\lambda},
\end{equation}
where $D$ is the largest physical linear dimension of the antenna and $\lambda$ is the wavelength.

The received power in dBm at distance $d$ is
\begin{equation}\label{mean_received_power}
\overline{P_r}(d) = P_t - \overline{PL}(d),
\end{equation}
where $P_t$ is transmit power in dBm.


\begin{table}[htbp!]
\centering
\caption[]{RSRP measurement report mapping in LTE}
\label{lte_rsrp}
\vspace{.2cm}
\begin{tabular}{cl}
\hline
Reported value & \begin{minipage}[t]{3cm} Measured value (dBm) \end{minipage} \\
\hline
\noalign{\smallskip}
RSRP$\_$00 & RSRP $<$ -140 \\
RSRP$\_$01 & -140 $\leq$ RSRP $<$ -139 \\
RSRP$\_$02 & -139 $\leq$ RSRP $<$ -138 \\
   ...  &\hspace{.6cm} ... \\
RSRP$\_$96 &  -45 $\leq$ RSRP$<$ -44 \\
RSRP$\_$97 &  -44 $\leq$ RSRP \\
\hline
\end{tabular}
\end{table}

\begin{figure*}[htbp!]
\centering
\includegraphics[width=\linewidth]{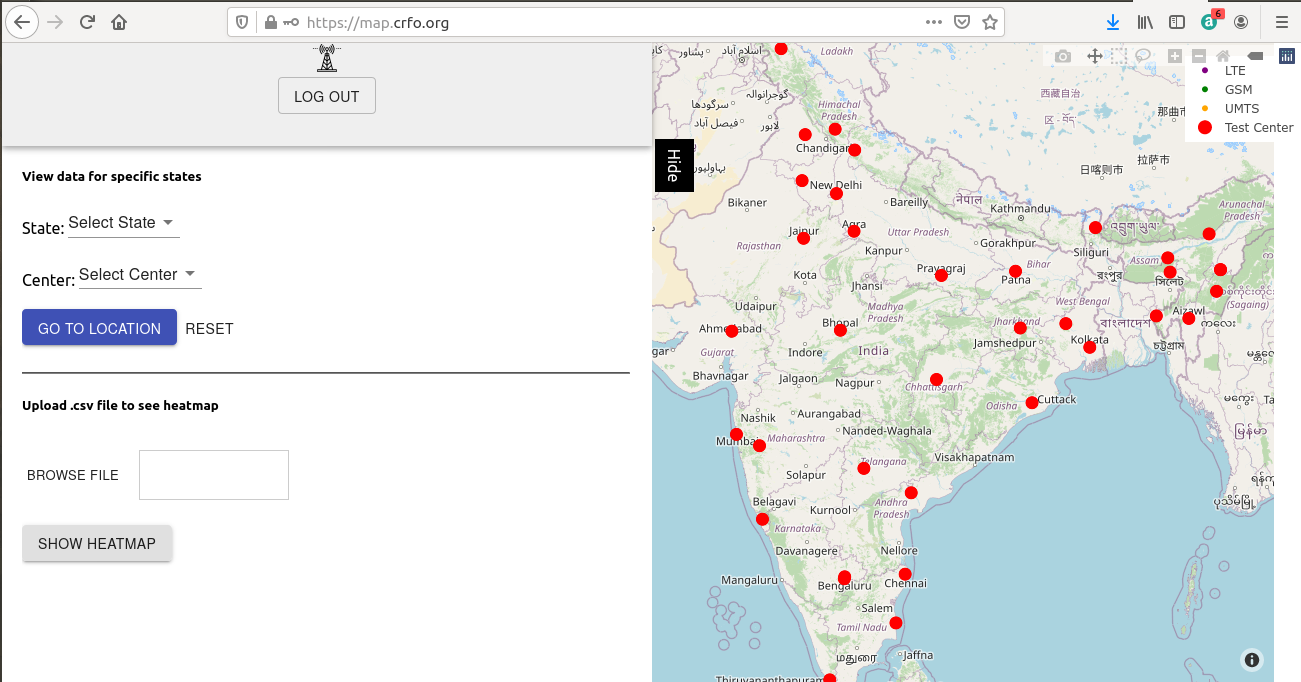}
\caption{CRFO web interface}
\label{india_map_crfo}
\end{figure*}

\subsection{Collecting RSRP samples using  TRAI MySpeed app}
In cellular communications, through the downlink transmissions from the base stations (BSs), the user equipment (UEs) or mobile phone periodically measures RSRP and other metrics (RSRQ and RSSI) for performing cell selection/re-selection and handover. For example, in LTE (4G) networks, there are 98 RSRP values defined, ranging from -140 dBm to -44 dBm with a 1 dB resolution as shown in Table \ref{lte_rsrp} \cite{rsrp_lte}.

The RSRP values can be collected using TRAI MySpeed app by following the below steps:
\begin{enumerate}
    \item Enable Mobile Internet
    \item Open TRAI MySpeed app.
    \item Press on `BEGIN TEST'. Wait for a few seconds to complete.
    \item Move to another location. Make sure that the distance between the current and next locations is greater than 20 m.
    \item Repeat Step 3.
\end{enumerate}

Figure \ref{TRAI app} shows the screenshots of TRAI MySpeed app as prescribed in above steps. Figure \ref{MySpeedTrai_result} shows the result of single measurement containing RSRP, latitude, longitude, download and upload speed. After conducting sufficient number of measurements, create a .CSV file in an appropriate format as explained in the next section.

\subsection{Generating CSV file}\label{csv_generation}
Create a .CSV file with the following three columns: `lon', `lat', and `val'. Fill the longitude and latitude values (with all decimal places) obtained from  TRAI MySpeed app in the first two columns of the .CSV file. The purpose for considering all decimal places of latitude and longitude value is to clearly distinguish the measurememt positions in the generated heatmap. The corresponding entries under `val' column can be either RSRP or Downlink or Uplink Speed.


\subsection{Procedure to view RSRP heatmap}\label{procedure_towers}

\begin{figure}[htbp!]
\centering
\includegraphics[width=\linewidth]{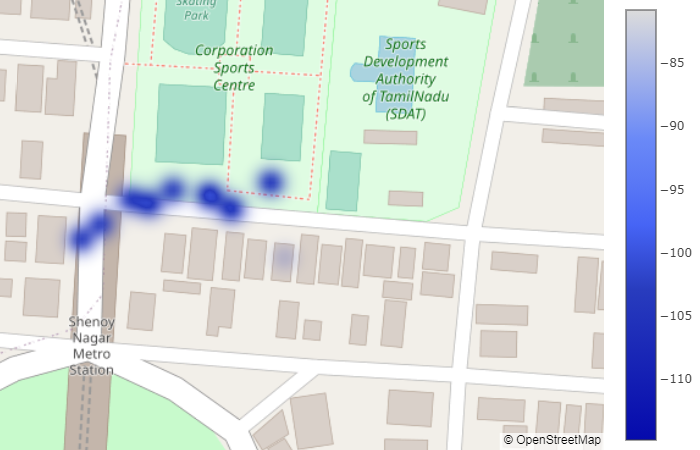}
\caption{CRFO Screenshot of Scenario 1: Near Chandhar Research Labs, Chennai}
\label{CRFO_screenshot1}
\end{figure}

\begin{figure}[!htbp]
\centering
\includegraphics[width=\linewidth]{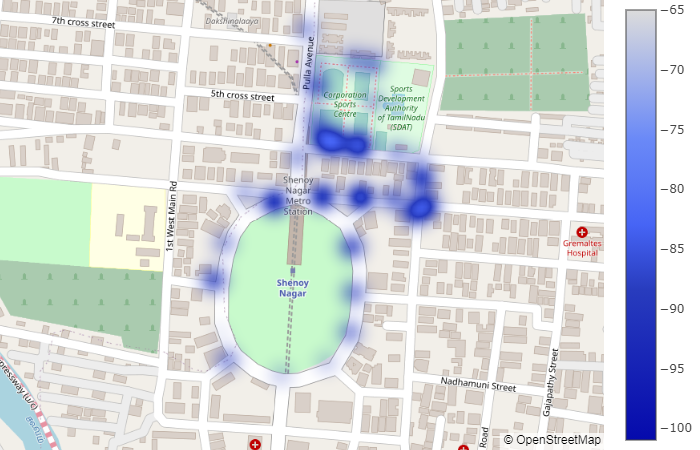}
\caption{CRFO Screenshot of Scenario 2: Near Chandhar Research Labs, Chennai}
\label{CRFO_screenshot2}
\end{figure}

\begin{figure}[!htbp]
\centering
\includegraphics[width=\linewidth]{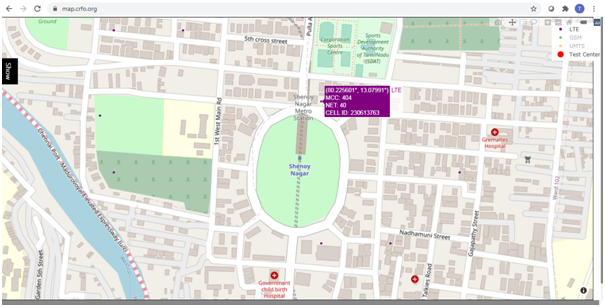}
\caption{Tower location around the measurement location: Near Chandhar Research Labs, Chennai}
\label{CRFO_Tower-screenshot3}
\end{figure}

\begin{enumerate}
	\item Open the website https://map.crfo.org in a web browser.
	\item Click on ``CRFO map" button (next to ``Home" button) in the CRFO webpage.
	\item Click on the link ``Welcome to CRFO map".
	\item Create a new account by registering with your email id.
	\item After successful registration, login to CRFO map with your email id and password. 
	\item Now, login into ``Welcome to CRFO map" link.
	\item The opened window is divided into two regions such as selection page on left side and Indian map on right side  as shown in Figure \ref{india_map_crfo}.
	\item Now, choose the tabs ``Select State" and ``Select Center" as per your choice and click on ``GO TO LOCATION" to view data for specific states and location.
	\item Now, click on ``BROWSE FILE" to upload your .csv file (See Section \ref{csv_generation}) and click on ``SHOW HEATMAP" to visualise your locations on the map page. You can zoom in/out the map and select/unselect LTE/GSM/UMTS/Test  for better visualization and understanding of generated map plot as shown in Figures \ref{CRFO_screenshot1} and \ref{CRFO_screenshot2}. (Note: All the icons are present on top right side Indian map page). 
	\item Now click on ``SHOW TOWERS" on selection page to visualize the towers present around your location/ test center/ generated map as shown in Figure \ref{CRFO_Tower-screenshot3}.
	\item Click on ``LOG OUT" to return home page.
\end{enumerate}

\subsection{Estimation of distance between UE and BS from RSRP samples}
In practice, the pathloss between the BS and UE depends on the scenario such as Urban Micro (UMi), Urban Macro (UMa), and Rural Macro (RMa) and Line-of-Sight (LoS) and Non-line of sight (NLoS) conditions \cite{m2135,m2412}.

The pathloss formula has to be selected based on the above mentioned scenarios. Since, our measurements are taken in Urban scenario (Chennai city), let us consider pathloss formula for NLoS condition in UMi scenario given by \cite[Tab.A1-2]{m2135}
\begin{equation}  \label{pathloss}
 PL(d) = 36.7\log d+22.7+26\log f_c,
 \end{equation}
where $f_c$ is carrier frequency in GHz and $d$ is distance in meters.

From \eqref{pathloss} and \eqref{mean_received_power}, the distance ($d$) in meters can be calculated as
\begin{equation}\label{d_est}
\hat{d} = 10^{\frac{PL(d)-22.7-26 \log f_c}{36.7}}.
\end{equation}


Tables \ref{table_scenario1} and \ref{table_scenario2} show the estimated distance from \eqref{d_est} for scenarios as shown in Figures \ref{CRFO_screenshot1} and \ref{CRFO_screenshot2}, respectively, with transmit power, $P_t$ = 41 dBm. The RSRP values are collected using the smartphone which is connected to Vodafone network with carrier frequency $f_c=$ 2320 MHz. The pathloss and the distance for each RSRP sample are calculated using Equations \ref{pathloss} and \ref {d_est}, respectively. From Tables \ref{table_scenario1} and \ref{table_scenario2}, it can be observed that the pathloss varies from 106 dB to 156 dB. The estimated distances can be verified from the maps as shown in Figures \ref{CRFO_screenshot1} and \ref{CRFO_screenshot2}. If the estimated distances are different from the actual distances, then the pathloss formula in Equation \eqref{pathloss} has to be changed according to \cite[Tab.A1-2]{m2135}. For example, the pathloss formula for LoS condition can be used if the measurement is taken in front of a tower (i.e. no building in between the tower and the measurement location).

\begin{table}[htbp!]
\centering
\footnotesize
\caption{Distance calculation from RSRP values with measurement scenario 1; $P_t$ = 41 dBm, $f_c$ = 2.32 GHz}\vspace{.2cm}
\label{table_scenario1}
\begin{tabular}{|l|l|c|c|c|}
\hline
\textbf{Latitude} & \textbf{Longitude} & \textbf{RSRP (dBm)} & \textbf{$P_L(d)$ (dB)} & \textbf{$\hat{d}$ (m)} \\ \hline
13.0801679        & 80.2260928         & -109        & 150                   & 1581          \\ \hline
13.0801522        & 80.2260598         & -115        & 156                   & 2332           \\ \hline
13.0804893        & 80.2260314         & -86        & 127                  & 376        \\ \hline
13.0803774        & 80.225859          & -84        & 125                 & 345           \\ \hline
13.080398         & 80.2254994         & -86        & 127                 & 388           \\ \hline
13.0804202        & 80.2254112         & -89           & 130                    & 462           \\ \hline
13.0802461        & 80.2251972         & -88        & 129                  & 428           \\ \hline
13.0803134        & 80.2252857         & -88        & 129                  & 437          \\ \hline
13.0804581        & 80.2256027         & -88        & 129                 & 442          \\ \hline
13.0804357        & 80.225761          & -81        & 122                 & 277          \\ \hline
\end{tabular}
\end{table}
\normalsize
 
\begin{table}[htbp!]
\centering
\footnotesize
\caption{Distance calculation from RSRP values with Measurement Scenario 2; $P_t$ = 41 dBm, $f_c$ = 2.32 GHz}\vspace{.2cm}
\label{table_scenario2}
\begin{tabular}{|l|l|c|c|c|}
\hline
\textbf{Latitude} & \textbf{Longitude} & \textbf{RSRP (dBm)} & \textbf{$P_L(d)$ (dB)} & \textbf{$\hat{d}$ (m)} \\ \hline
13.0801464        & 80.2260452         & -101          & 142                     & 981           \\ \hline
13.0804985        & 80.2257557         & -75           & 116                     & 192           \\ \hline
13.0805631        & 80.2255582         & -79           & 120                     & 247           \\ \hline
13.0808002        & 80.2254647         & -87           & 128                     & 408           \\ \hline
13.0812191        & 80.2254888         & -79           & 120                     & 247           \\ \hline
13.0819614        & 80.225355          & -89           & 130                     & 462           \\ \hline
13.0815274        & 80.2254982         & -91           & 132                     & 524           \\ \hline
13.0815553        & 80.225979          & -87           & 128                     & 408           \\ \hline
13.0815046        & 80.226252          & -85           & 126                     & 360           \\ \hline
13.0804931        & 80.2260634         & -79           & 120                     & 247           \\ \hline
13.0800177        & 80.2268996         & -73           & 114                     & 170           \\ \hline
13.0802885        & 80.2268616         & -95           & 136                     & 673           \\ \hline
13.0796643        & 80.2269571         & -71           & 112                     & 149           \\ \hline
13.080351         & 80.2265585         & -89           & 130                     & 462           \\ \hline
13.0795623        & 80.2267914         & -77           & 118                     & 218           \\ \hline
13.0797573        & 80.2265138         & -83           & 124                     & 317           \\ \hline
13.0797832        & 80.2255705         & -69           & 110                     & 132           \\ \hline
13.0797705        & 80.2260777         & -65           & 106                     & 103           \\ \hline
13.0780113        & 80.2259357         & -87           & 128                     & 408           \\ \hline
13.0785264        & 80.225988          & -85           & 126                     & 360           \\ \hline
13.0776225        & 80.225624          & -95           & 136                     & 673           \\ \hline
13.0774578        & 80.2250176         & -95           & 136                     & 673           \\ \hline
13.0780228        & 80.2241998         & -99           & 140                     & 866           \\ \hline
13.0776462        & 80.224294          & -87           & 128                     & 408           \\ \hline
13.0786936        & 80.2241142         & -79           & 120                     & 247           \\ \hline
13.0791813        & 80.2242534         & -89           & 130                     & 462           \\ \hline
13.0797125        & 80.2249236         & -69           & 110                     & 132           \\ \hline
13.0798296        & 80.2245243         & -91           & 132                     & 524           \\ \hline
13.080434         & 80.2256389         & -89           & 130                     & 462           \\ \hline
13.0802891        & 80.2252336         & -89           & 130                     & 462           \\ \hline
13.0803889        & 80.226053          & -83           & 124                     & 317           \\ \hline
13.0810215        & 80.2262463         & -91           & 132                     & 524           \\ \hline
13.0791252        & 80.2259602         & -79           & 120                     & 247           \\ \hline
\end{tabular}
\end{table}
\normalsize

To know the exact LTE tower locations follow the instructions given in Section \ref{procedure_towers}. After step 10, you can choose to view LTE cell towers by clicking toggle buttons on top right corner of the page (See Figure \ref{india_map_crfo}). Note that the tower positions shown in Figure \ref{CRFO_Tower-screenshot3} are approximate positions. Therefore, we have to physically verify the location by visiting the tower locations. After finding the tower locations, we can verify the distances obtained using Equation \eqref{d_est}.

\section{Experiment 2: Verification of Shadow fading distribution}

The objective of this experiment is to verify the Shadow fading  model from the real time signals received from the nearby 4G cellular BSs.

\begin{figure*}[htbp!]
\centering
\subfigure[]{\includegraphics[scale=.165]{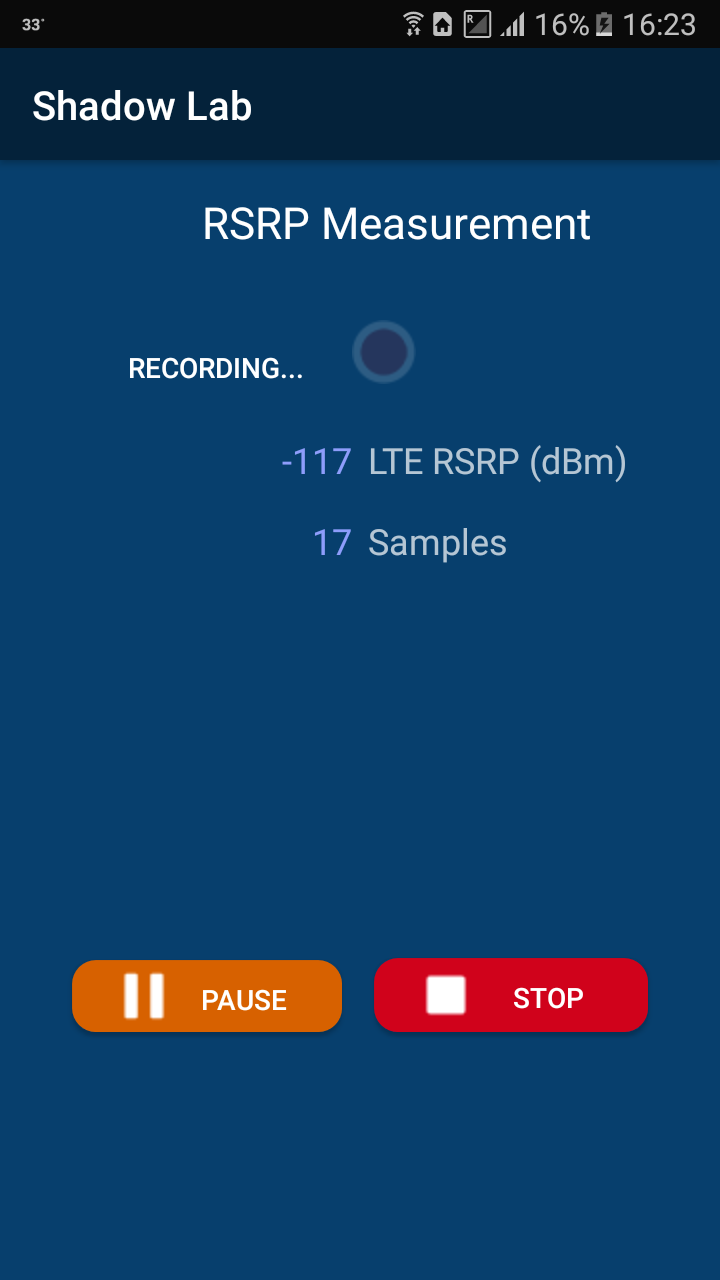}\label{shadow_lab_recording2}}
\subfigure[]{\includegraphics[scale=.11]{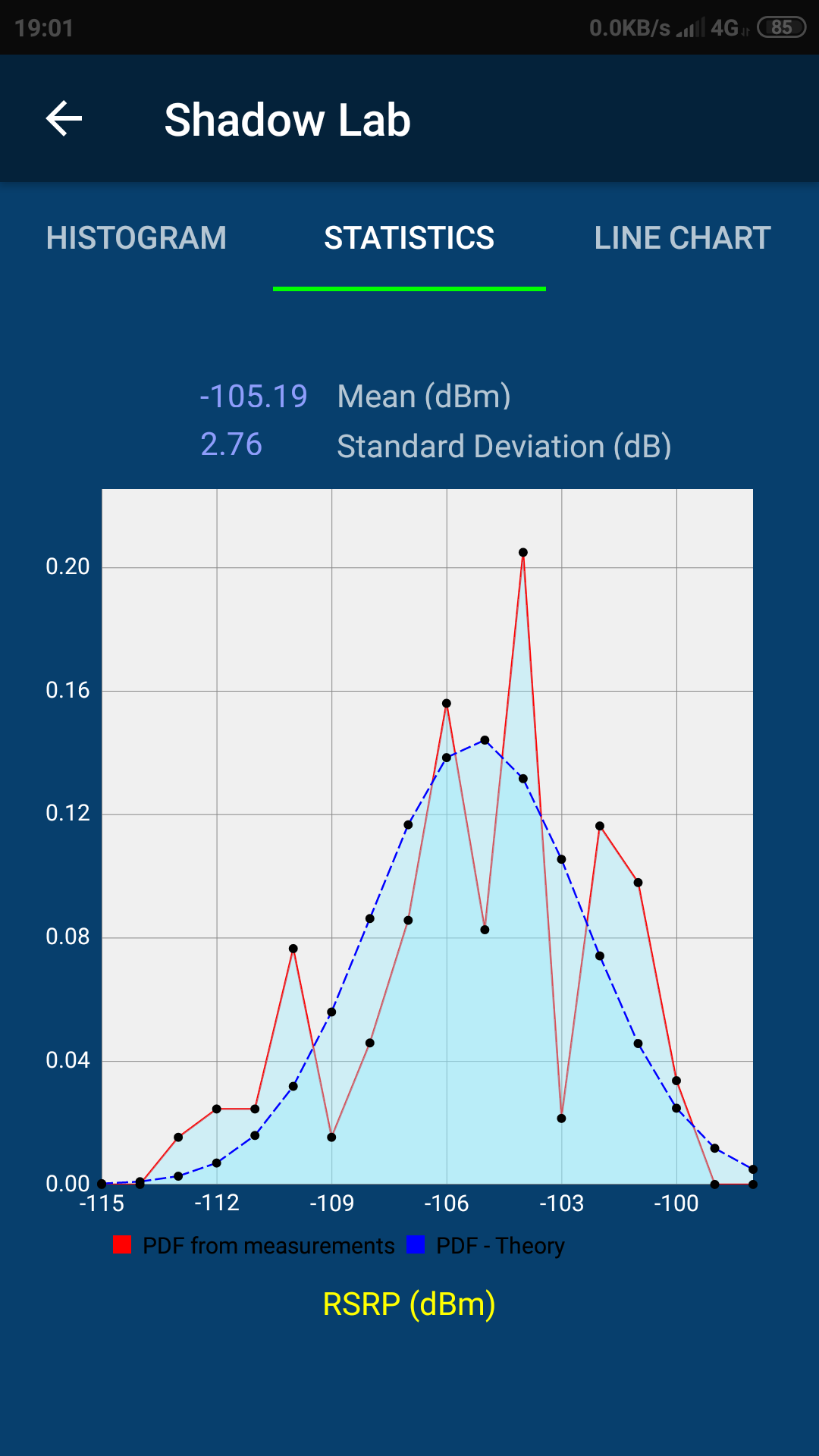} \label{shadow_lab_res1}}
\subfigure[]{\includegraphics[scale=.11]{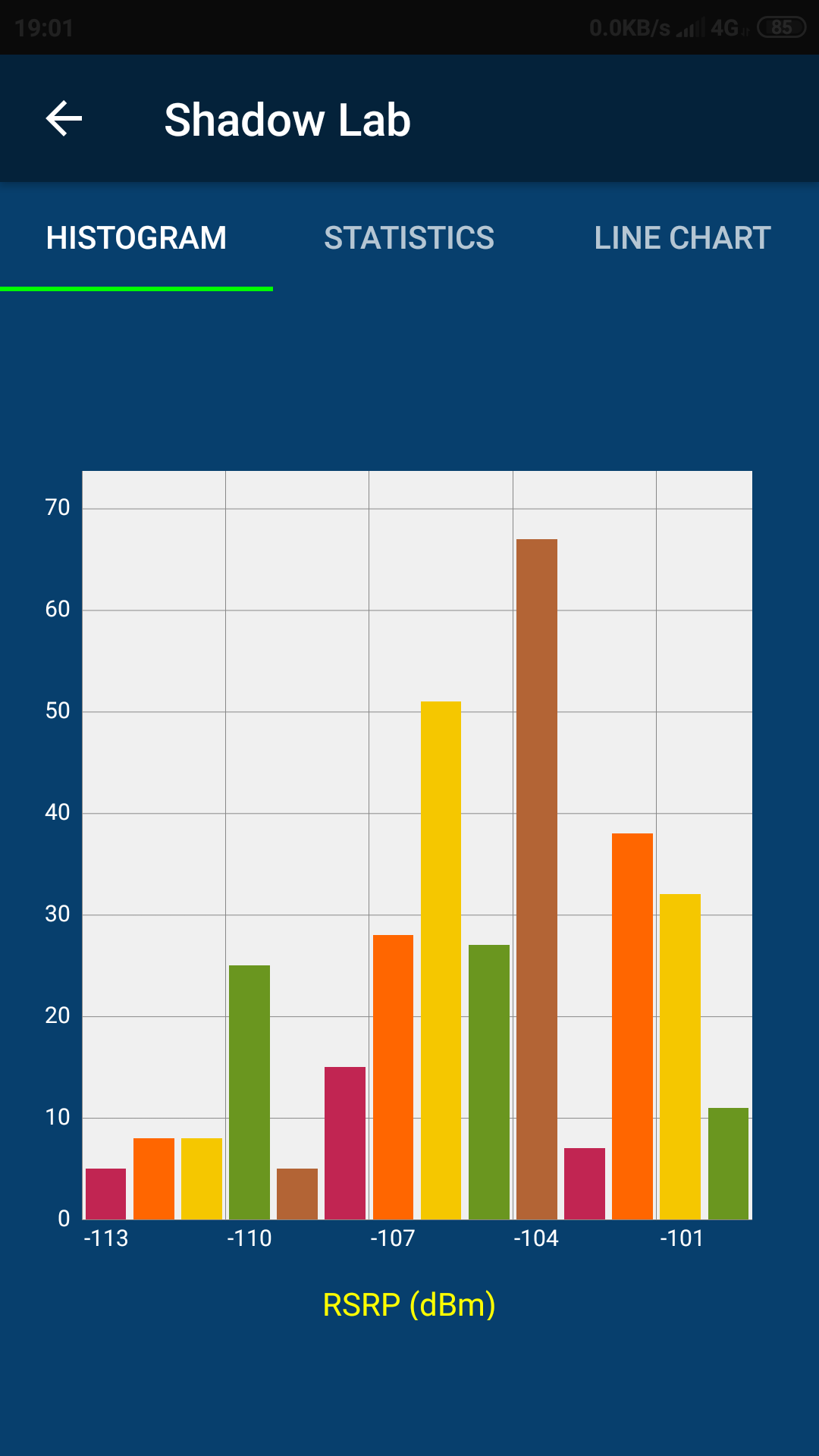}\label{shadow_lab_res2}}
\subfigure[]{\includegraphics[scale=.11]{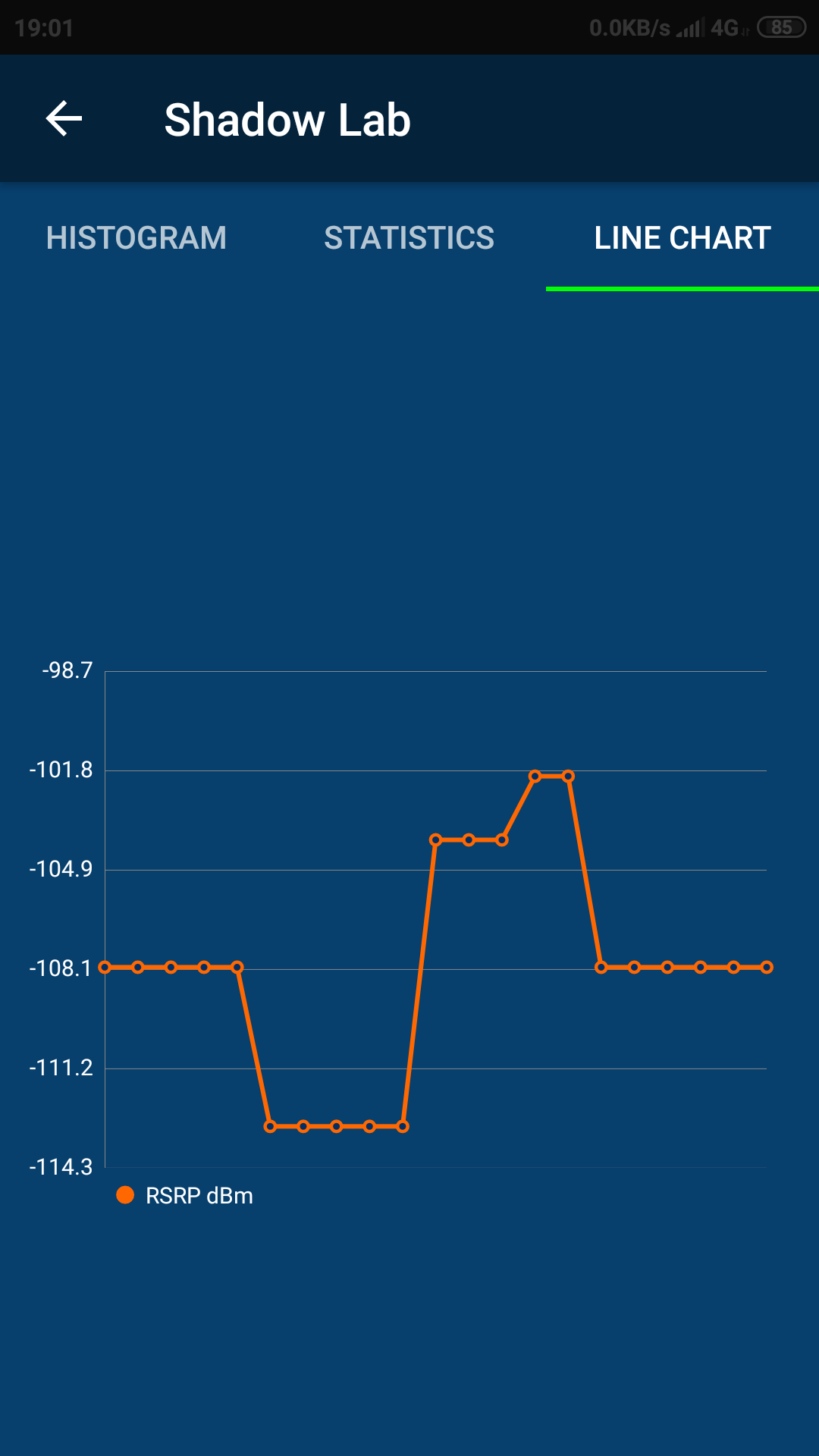}\label{shadow_lab_res3}}
\caption{Screenshots of `Shadow Lab' app}
\label{shadow_lab_screenshot}
\end{figure*}

\subsection{Shadow fading}
Shadow fading (or Shadowing) plays an important role in cellular network planning and operation as it significantly affects the coverage performance. As a result of shadow fading, for the same transmitter-receiver separation, the received signal strength
may be vastly different.

If the received signal strength is measured in dB units, then the shadow fading is modeled as zero-mean Gaussian distributed random variable with standard deviation (in dB). For detailed understanding of shadow fading, you may refer to \cite[Sec:4.9.2]{Rappaport2001}. 

Due to the objects in surrounding environment, the received signal strength at the receiver fluctuates around the mean value at distance $d$ as given in \eqref{mean_received_power} is  distributed normally. That is
\begin{equation}
P_r(d) = \overline{P_r}(d)+\xi \ \ \ \text{dBm},
\end{equation}
where $\xi$ is zero-mean Gaussian distributed RV in dB with standard deviation $\sigma_\xi$. In practice, the standard deviation typically varies from 4 dB (Urban cellular) to 12 dB (Rural)  \cite[Tab:A1-2]{m2135}.

The PDF of the normally distributed received power $P_{r}$ (in dBm) is given as 
\begin{equation}\label{pdf_normal}
f_{P_{r}}(x)=\frac{1}{ \sigma_{P_{r}}\sqrt{2\pi}} \exp{\frac{-(x-\mu_{P_{r}})^2}{2\sigma_{P_{r}}^2}},
\end{equation}
where $\mu_{P_r} = \overline{P_r}$ and $\sigma_{P_r} = \sigma_\xi$.



\subsection{Collecting LTE RSRP samples using android based smart phone}

The `Shadow Lab' app collects the RSRP samples at regular time intervals (currently for every second) and stores them in a file in .csv format.

Measurements can be taken inside the building or across the street. After clicking `New Recording' button, slowly start walking. Now RSRP values be displayed as shown in Figure \ref{shadow_lab_recording2}. Take a step and stand for 5 to 10 seconds. Take another step and stand for 5 to 10 seconds. Continue this process until you covered the entire area of your building or street.

If you are inside home or office, try to take measurements in all rooms. If you are in outdoor areas, take measurements in a circular area with 5 m radial distance. If sufficient number of samples are collected (more than 500), then press the stop button. Now a csv file will be generated in the directory: Device Storage $\rightarrow$ Data $\rightarrow$ Android $\rightarrow$ Shadow Lab. 


\subsubsection{Estimation of mean and variance}
Let $N$ be the total number of collected samples. The mean and variance are estimated as follows.

Mean:
\begin{equation}
\hat{\mu}_{P_r} =  \frac{1}{N}\sum_{n=1}^N x_n
\end{equation}
Variance:
\begin{equation}
 \hat{\sigma}_{P_r}^2 = \frac{1}{N-1}\sum_{n=1}^N (x_n-\hat{\mu}_{P_r})^2
\end{equation}

Figure \ref{shadow_lab_res1} shows the comparison between the PDF of measured RSRP samples and theory with three different mean RSRP values. The obtained statistical values with its histogram plot is shown in Figures \ref{shadow_lab_res2} and \ref{shadow_lab_res3}, respectively.

\section{Experiment 3: Verification of small scale fading distribution}

The objective of this experiment is to verify the small scale fading model from the real time signals received from the nearby cellular BSs. 


\subsection{Small scale fading}
Small scale fading models are used to describe the signal fluctuations created by multi-path components over short period of time or travel distance. The objects and scatterers in the environment leads to multiple versions of the signal energy at the receiving antenna. Due to multi-path effects, the signal energy arrives at the receiver with random phase and time lag depending on the path between the transmitter and receiver. Multi-path propagation creates unwanted effects called inter-symbol-interference (ISI) in which one symbol interferes with subsequent symbols. Doppler shift is another issue created by multi-path propagation. 

Let $n$-th reflected wave with amplitude $a_n$ arrival from an angle $\theta_n$ relative to the direction of motion of the antenna. Then the Doppler shift due to user mobility is given by \cite{Rappaport2001}
\begin{equation}\label{doppler_shift}
 f_{D,n} = \frac{v}{\lambda} \operatorname{cos} \theta_n \ \text{Hz},
\end{equation}
where $v$ is velocity of the user and $\theta_n$ is an angle between $n$-th reflected wave and direction of the receiver. It can be observed from Equation \eqref{doppler_shift} that the Doppler shift is positive when the user is moving towards the direction of the arrival of the wave and it is negative if the user is moving away from the direction of the arrival of the wave. 

The maximum Doppler shift occurs when the user is moving opposite to the direction of the incoming wave i.e.
\begin{equation}
f_m = \frac{v}{c} f_c,
\end{equation}
where $f_c$ is carrier frequency and $c$ is velocity of light.

Consider the transmitted band-pass signal
\begin{equation}
 s(t) = \mathrm{Re}\big[\tilde{s}(t)\ e^{j2\pi f_ct}\big],
\end{equation}
where $\tilde{s}(t)$ is the complex envelope of the transmitted signal and $\mathrm{Re}(x)$ denotes the real part of $x$. If the band-pass signal pass through the channel which comprises $N$ multipath components, then the noiseless received band-pass signal is \cite{Stuber2001,Rappaport2001}
\begin{equation}\label{received_multipath}
 r(t) = \mathrm{Re}\bigg[\sum_{n=1}^{N}a_ne^{j2\pi[(f_c+f_{D,n})(t-\tau_n)]}\tilde{s}(t-\tau_n)\bigg],
\end{equation}
where $a_n$ and $\tau_n$ are the amplitude and time delay, respectively, of the $n$-th path. Here $a_n$ depends on the dimensions of the reflecting surface and diffracting edge.

The received band-pass signal in Equation \eqref{received_multipath} can be rewritten as
\begin{equation}
 r(t) = \mathrm{Re}\big[\tilde{r}(t)\ e^{j2\pi f_ct}\big],
\end{equation}
where the received complex envelope is 
\begin{equation}
 \tilde{r}(t) = \sum_{n=1}^{N}a_n\ e^{-j\phi(t)}\tilde{s}(t-\tau_n),
\end{equation}
where $\phi_n(t) = 2\pi\big[(f_c+f_{D,n})\tau_n-f_{D,n}t\big]$.

\paragraph{Received Envelope Distribution:}
The received band-pass signal can be further rewritten as
\begin{equation}
r(t) = g_I(t)\operatorname{cos}(2\pi f_c t) - g_Q(t) \operatorname{sin}(2\pi f_c t),
\end{equation}
where \begin{equation}
       g_I(t) = \sum_{n=1}^{N}a_n \operatorname{cos} \phi_n(t)
      \end{equation}
and
\begin{equation}
       g_Q(t) = \sum_{n=1}^{N}a_n \operatorname{sin} \phi_n(t)
      \end{equation}
are the in-phase and quadrature phase components of the received band-pass signal. For large $N$, by using central limit theorem $g_I(t)$ and $g_Q(t)$ can be treated as Gaussian random processes. Both $g_I(t)$ and $g_Q(t)$ are uncorrelated zero-mean Gaussian RVs with an equal variance given by $\overline{g_I^2} = \overline{g_Q^2} = b_0$. The total received envelope power is expressed as $\sigma = \overline{g_I^2} +\overline{g_Q^2} = \sum_{n=1}^{N}a_n^2 = 2b_0$.

The envelope of the received band-pass signal i.e. $|r(t)|$= $\sqrt{g_I^2(t)+g_Q^2(t)}$ has a Rayleigh distribution given by \cite{Stuber2001}
\begin{equation}\label{pdf_eqn}
 p_{|r(t)|}(x) = \frac{2x}{\sigma} \exp{\bigg(-\frac{x^2}{\sigma}\bigg)},\ x\geq0.
\end{equation}
\subsection{Sample collection using Wi-Guy\textsuperscript{\textregistered}}
 
The baseband samples are collected using an SDR based device `Wi-Guy\textsuperscript{\textregistered}' \cite{wi-guy}, a low cost platform for understanding advanced wireless communication technologies. The Wi-Guy\textsuperscript{\textregistered} converts real-time continuous signals captured by the antenna into discrete time baseband signals and stores it in a `.dat' file\footnotemark. 

\footnotetext{The .dat files are available in CRFO for offline processing.}

For processing, the samples in `filename.dat' file are converted into I-Q samples using a Python script\footnotemark. 

\footnotetext{Python script is available in CRFO}.

\begin{figure}[htbp!]
\centering
\vspace{-.25cm}
\includegraphics[width=\linewidth]{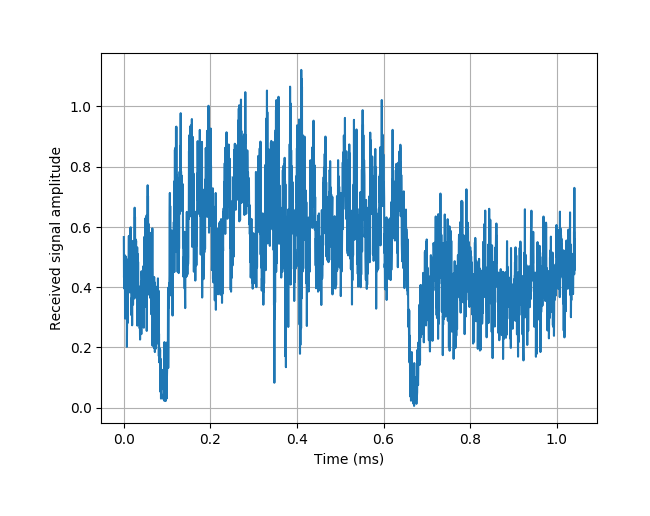}\vspace{-.5cm}
\caption{Amplitude of the signal captured at 938.8 MHz}
\label{Exp1_Received_Signal}
\end{figure}
 
In order to verify the distribution, it is important to select a suitable frequency which has active transmission. In this experiment, the frequency 938.8 MHz is chosen because it is an active channel in the GSM900 band\footnotemark. The absolute values of the received samples are plotted as shown in Figure \ref{Exp1_Received_Signal}. Figure \ref{Exp1_Received_Signal} shows the received signal amplitude over the period of 1 ms. It can be observed that transmission is discontinuous and nulls appear in between the transmissions. It is clearly seen that the difference between the two nulls is exactly equal to 576.9 $\mu$s, which is the slot duration in GSM cellular system.
\footnotetext{Video demonstration on finding active GSM channels is available in this link: \url{https://youtu.be/sBo69Ufo0Y4}}

From the absolute values of the received samples, the scale parameter is estimated as 
\begin{equation}
 \hat{\sigma} =  \sqrt{\frac{1}{2N}\sum_{i=1}^N r_i^2}.
\end{equation}

Then using `rayleigh' function, the theoretical PDF is generated.

\begin{figure}[htbp!]
\centering
\includegraphics[width=\linewidth]{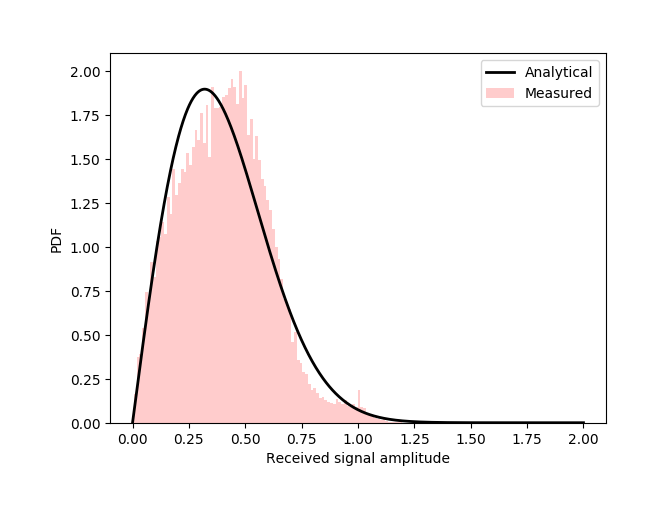}
\vspace{-.5cm}
\caption{PDF of received signal amplitude}
\label{Exp1_Rayleigh_PDF}
\end{figure}

Figure \ref{Exp1_Rayleigh_PDF} shows the comparison between the PDFs obtained from the measured samples and using Equation \eqref{pdf_eqn}. It can be observed that the PDF obtained from the measurement samples closely matches with the theoretical PDF.

\section{Video demonstration}
Video demonstration of the experiments detailed in tutorial is available in this link: \url{https://youtu.be/pXxAzpmIixI}

\section{Summary}
In this tutorial, we detailed three simple experiments to understand the fundamentals of mobile communication concepts: pathloss, Shadow fading  and small scale fading using smartphone and CRFO without involving any expensive hardware. We hope that, in future, many new experiments will be developed using CRFO to further explore and develop future mobile communication systems.






\end{document}